\documentclass[twocolumn,superscriptaddress]{revtex4-2} 
\usepackage[T1]{fontenc}
\usepackage{times}
\usepackage{graphicx}
\usepackage{booktabs}

\usepackage{amsfonts, amsmath,amssymb, bm, graphicx}
\usepackage{siunitx}

\usepackage{amssymb}
\usepackage[english]{babel}
\usepackage{lipsum}

\newcommand{\figref}[2]{\hyperref[#1]{\ref{#1}(#2)}}
\newcommand{\figrefsub}[3]{\hyperref[#1]{\ref{#1}(#2)#3}}

\usepackage{physics}
\usepackage{lipsum}
\usepackage{changes}

\usepackage[colorlinks=true,allcolors=blue]{hyperref}
\usepackage{footmisc}

\usepackage{upgreek}

\usepackage{soul}

\makeatletter
\let\ORIbbl@fixname\bbl@fixname
\def\bbl@fixname#1{%
  \@ifundefined{languagealias@\expandafter\string#1}
    {\ORIbbl@fixname#1}
    {\edef\languagename{\@nameuse{languagealias@#1}}}%
}
\newcommand{\definelanguagealias}[2]{%
  \@namedef{languagealias@#1}{#2}%
}
\makeatother

\definelanguagealias{en}{english}

\begin{document}

{
\makeatletter
\def\frontmatter@thefootnote{%
 \altaffilletter@sw{\@fnsymbol}{\@fnsymbol}{\csname c@\@mpfn\endcsname}%
}%

\makeatother

\title{Mode splitting of spin waves in magnetic nanotubes with discrete symmetries}

\author{Lukas K\"orber}\email{l.koerber@hzdr.de}
\affiliation{Helmholtz-Zentrum Dresden - Rossendorf, Institut f\"ur Ionenstrahlphysik und Materialforschung, D-01328 Dresden, Germany}
\affiliation{Fakultät Physik, Technische Universit\"at Dresden, D-01062 Dresden, Germany}

\author{Istv\'{a}n K\'{e}zsm\'{a}rki}
\affiliation{Experimental Physics V, University of Augsburg, 86135 Augsburg, Germany}

\author{Attila K\'{a}kay}
\affiliation{Helmholtz-Zentrum Dresden - Rossendorf, Institut f\"ur Ionenstrahlphysik und Materialforschung, D-01328 Dresden, Germany}

\date{\today}

\begin{abstract}
We investigate how geometry influences spin dynamics in polygonal magnetic nanotubes. We find that lowering the rotational symmetry of nanotubes by decreasing the number of planar facets, splits an increasing number of spin-wave modes, which are doubly degenerate in cylindrical tubes. This symmetry-governed splitting is distinct form the topological one recently observed in cylindrical nanotubes. Doublet modes with half-integer or integer multiple azimuthal periods of the number of facets, split to singlet pairs with lateral standing-wave profiles of opposing mirror-plane symmetries. Furthermore, the polygonal geometry facilitates the hybridization of modes with different azimuthal periods but the same symmetry, manifested in avoided level crossings. These phenomena, unimaginable in cylindrical geometry, provide new tools to control spin dynamics on nanoscale. The presented concept can be generalized to nano-objects of versatile geometries and order parameters, offering new routes to engineer dynamic response in nanoscale devices.
\end{abstract}

\maketitle

The field of magnetism provides many interesting examples of mesoscopic effects which emerge when transitioning from flat to curved or buckled geometries \cite{Smith11,Volkov19c,Volkov19,Streubel12,Streubel13c}. For example, the surface curvature of thin magnetic membranes can be used to stabilize nontrivial magnetic textures such as skyrmions  \cite{KravchukPRB2016SphericalShell,elias_winding_2019} which, in flat systems, are only stable in the presence of intrinsic Dzyaloshinskii-Moriya interactions or frustration. The spin-Cherenkov effect \cite{Yan11a,Yan13}, that emerges in magnetic nanotubes when domain walls propagate faster the spin-wave velocity, is another phenomenon not observed in flat geometries so far.
The effect of curvature on magnetization dynamics manifests itself, for example, in a nonreciprocal propagation of spin waves along magnetic nanotubes in the flux-closure (vortex) state [see Fig.~\figrefsub{fig:FIG1}{a}]. In this case, the surface curvature leads to an asymmetric spin-wave dispersion, i.e. waves counter-propagating along the nanotube exhibit different oscillation frequencies \cite{otaloraCurvatureInducedAsymmetricSpinWave2016}. These phenomena are induced by emergent magneto-chiral interactions \cite{bordacs_chirality_2012, kezsmarki_one-way_2014,GaidideiPRL_curvature_effects_2014,Sheka_2015,10.21468/SciPostPhys.10.3.078,sheka_nonlocal_2020} which, in return, stem from an inversion-symmetry breaking \cite{Szaller2013_Symmetry,Udvardi2009_chiral_asymmetry} by non-collinear spin textures embedded in curved magnetic membranes. 

Another important aspect of magnetism in specimen with complex three-dimensional shape is their topology. For example, in magnetic samples with a non-orientable surface, such as Möbius ribbons, topological magnetic solitons appear which are strongly tied to the geometry of the magnetic object~\cite{Pylypovskyi15b}. In magnetic rings and nanotubes, the topology of the magnetic waveguide can induce a Berry phase for the spin waves~\cite{dugaevBerryPhaseMagnons2005,salazar-cardonaNonreciprocitySpinWaves2021}. The spatial profile of the spin-wave modes in thin cylindrical magnetic nanotubes are described by $\exp[i(\nu\phi + kz)]$, where $\nu\in\mathbb{Z}$ counts the periods in azimuthal $(\phi)$ direction and $k\in\mathbb{R}$ denotes the wave-vector component along the axis ($z$ direction) of the nanotube. In the vortex equilibrium state, modes which are propagating clockwise ($\nu>0$) or counter clockwise ($\nu<0$) around the circumference of the tube form doublets. This azimuthal degeneracy is lifted for each $\nu$ by a Berry phase, when the magnetization is tilted from the vortex state to a helical state \cite{dugaevBerryPhaseMagnons2005,salazar-cardonaNonreciprocitySpinWaves2021}, as depicted in Fig.~\figref{fig:FIG1}{a}. Consequently, modes propagating in opposite azimuthal direction are gradually split into two lateral-running waves of different frequency with increasing axial component of the magnetization, as seen in Fig.~\figref{fig:FIG1}{b}. This topological effect is a nontrivial spin version of the Aharonov-Bohm effect~\cite{AharonovBohm_PRL1959,Mode_split_Ivanon_2005}, originally observed for charged particles in similar geometries, e.g., in carbon nanotubes \cite{bachtold_aharonovbohm_1999}.

\begin{figure}[h!]
    \centering
    \includegraphics{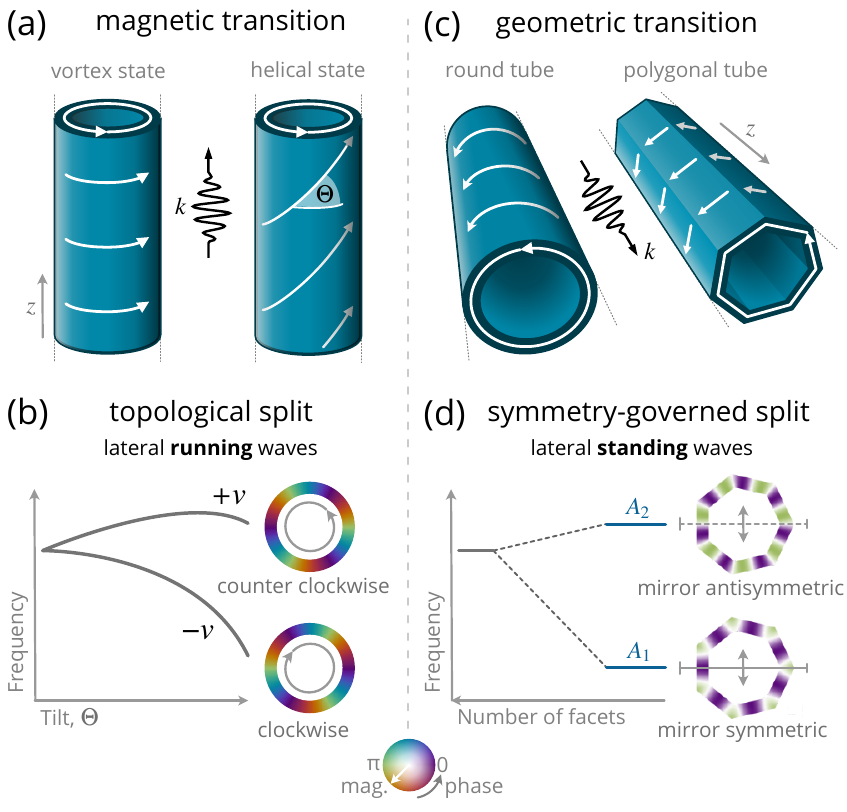}
    \caption{Comparison of the known topological doublet split with the newly presented symmetry-governed doublet split of spin waves in tubular magnetic waveguides. (a) Magnetic transition from vortex to axial state. In round vortex-state magnetic nanotubes counter-clockwise and clockwise modes (azimuthal index $\pm \nu$) are degenerate. (b) A topological split of these doublets appears when transitioning to the helical equilibrium state, which produces lateral running waves that are singlets. In contrast to this, (c) when the rotational symmetry of the tube is lowered, the symmetry-governed split of certain azimuthal doublets appears for tubes in the vortex state, leading to lateral standing waves with opposing mirror symmetry (here belonging to the $A_1$ and $A_2$ irreducible representations). (d) Splitting and lateral mode profiles of the spin waves using a bi-variate colormap, that displays the $z$ components of the oscillation magnitude and phase at the same time.}
    \label{fig:FIG1}
\end{figure}

Here, we explore another fundamental aspect of three-dimensional magnetic nano-objects, complementary to their topology, namely their symmetry. In particular, we investigate the evolution of the spin-wave spectrum in magnetic nanotubes when the continuous rotational symmetry of cylindrical tubes is lowered to a discrete rotational symmetry [Fig.~\figref{fig:FIG1}{c}]. In contrast to the topological doublet split and as a completely distinct feature, we observe the splitting of certain $\pm \nu$ doublet modes into singlets, for tubes in a vortex-like magnetic state. A symmetry classification of the modes according to the irreducible representations (irreps) belonging to the magnetic point group of the respective tube reveals that these singlet pairs have opposing mirror symmetry. For example, Fig.~\figref{fig:FIG1}{d} shows how the $\nu=\pm7$ doublet of a round tube is split  in a heptagonal tube into an $A_1$--$A_2$ singlet pair of lateral-standing-wave character. This geometric effect is distinct from the topological doublet split and independent of the involved magnetic interactions.


Our numerical study of the magnetization dynamics is carried out within the continuum limit of micromagnetism \cite{brownjr.Micromagnetics1963, gurevichMagnetizationOscillationsWaves1996, stancilSpinWavesTheory2009} in combination with group theory. At first, we introduce the geometry and the equilibrium magnetic state of the studied systems, which together define their magnetic symmetry groups. We model polygonal tubes with facets between $c=6$ and $c=30$ facets, and a round tube ($c=\infty$), as the limiting case. The magnetization in the equilibrium state, $\bm{m}_0$, is obtained for each tube by minimizing the total energy \cite{korberFiniteelementDynamicmatrixApproach2021}. The vortex states is stabilized by an easy-plane magneto-crystalline anisotropy, favoring an orientation of the magnetization in the plane of the cross section. In real samples, such an anisotropy can be achieved using epitaxial crystal growth \cite{zimmermannOriginManipulationStable2018}. Finally, the lateral mode profiles are obtained using our recently developed finite-element dynamic-matrix approach for propagating spin waves \cite{korberFiniteelementDynamicmatrixApproach2021}. Details of the numerical calculations are also found in Suppl. material S1. For nanotubes with a thin enough magnetic shell, $\bm{m}_0$ can be assumed to follow the geometric shape of the cross section: the magnetization is nearly homogeneous in the flat facets, whereas in the corners it forms domain walls [see Fig.~\figref{fig:FIG2}{a}].

Disregarding discretization effects, the symmetry group of the resulting flux-closure state is that of the idealistic polygonal vortex state with \textit{sharp corners}, which is $c/m^\prime m m$ when $c$ is even, and $c/m^\prime m$ when $c$ is odd. The odd-facet tubes have only one set of mirror planes connecting corners with facets, while even-facet tubes have one set connecting opposite corners and another set connecting opposite facets [see Fig.~\figref{fig:FIG2}{b}]. When transitioning from a cylindrical to a polygonal tube, the modes can be classified according to the irreps of $c/m^\prime m(m)$, which are either one or two dimensional~\cite{mprime_m_m}(see Table I. in Suppl. material S2). As a consequence, only singlet or doublet solutions are allowed in the spin-wave spectrum. The expected mode profiles in the phase space of the singlets (standing waves) and the doublets (running waves) are shown in Fig.~\figref{fig:FIG2}{c}. Note, that in contrast to the phase map of singlets, which are invariant upon all symmetry operations (apart from a factor of $-1$ for certain symmetry operations), the phase maps of doublet pairs are exchanged by some of the symmetries, again emphasizing the standing/running wave distinction. For round tubes there is only one singlet solution, the homogeneous mode, which for polygonal tubes, maintains a homogeneous phase profile within the cross section, but its oscillation magnitude is enhanced around the corners of the tube, as shown in Fig.~\figref{fig:FIG2}{d}.

\begin{figure}[h!]
    \centering
    \includegraphics{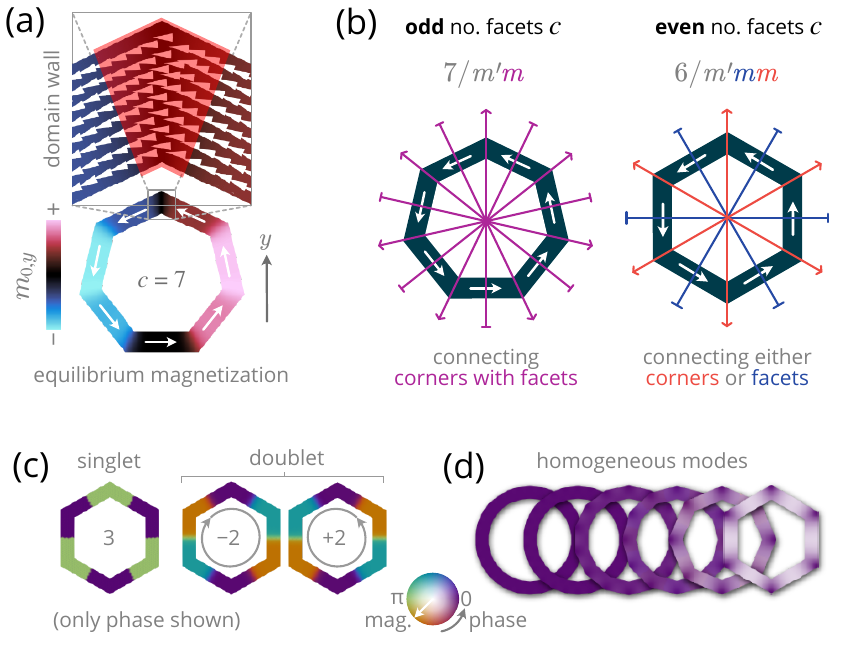}
    \caption{(a) Vortex equilibrium state in a heptagonal nanotube obtained from numerically simulations for a single cross section. The zoom in shows the presence of narrow domain walls in the corners where the magnetization rotates continuously between two neighboring facets. (b) Axial mirror planes (planes containing the long axis of the tube) for an odd- and an even-numbered polygonal tube. The odd-facet cross sections contain only one set of mirror planes (corner-to-facet), while even-facet ones contain two sets (corner-to-corner and facet-to-facet). (c) Mode profiles in the phase space of the standing-wave singlets and the running-wave doublets. (d) The transformation of the homogeneous lateral mode ($\nu=0$) into a mode localized to the corners of the polygonal cross section is summarized. For both magnitude and phase of the modes, only the $z$ component is shown.}
    \label{fig:FIG2}
\end{figure}


\begin{figure*}[t!]
    \centering
    \includegraphics{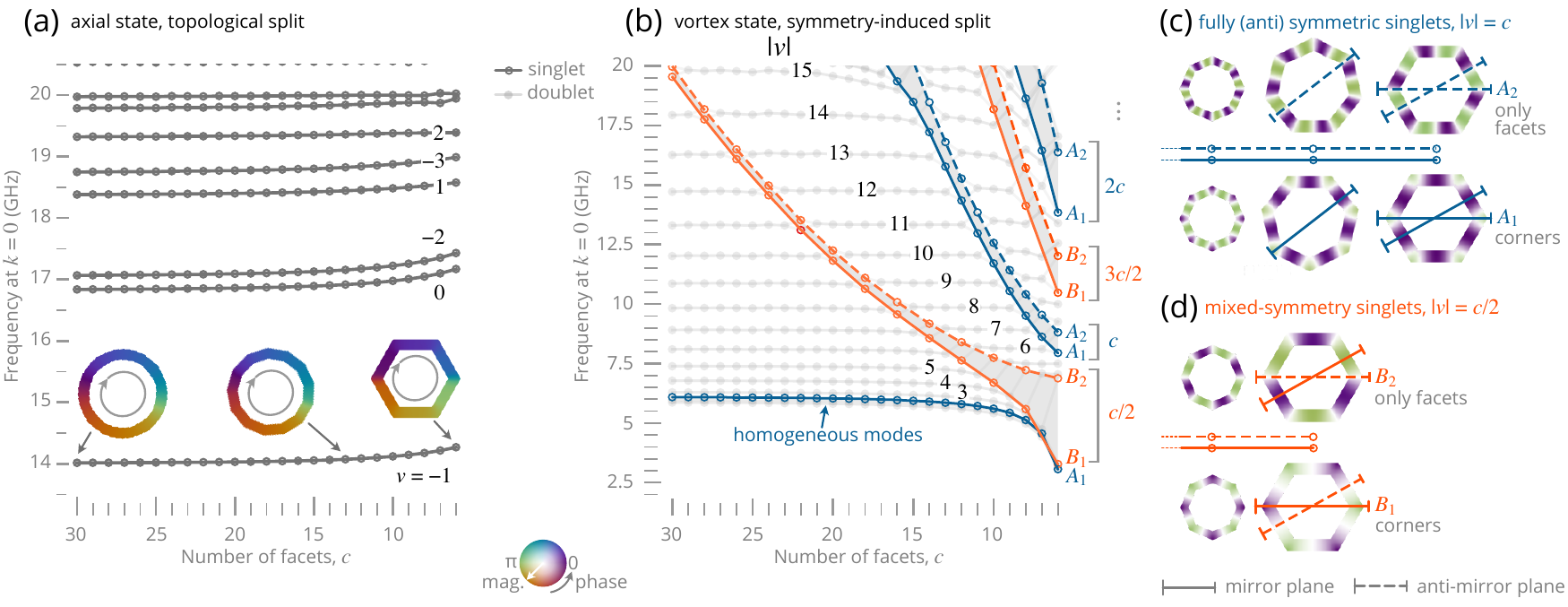}
    \caption{Frequencies of the spin waves at $k=0$ in nanotubes when the number of facets $c$ is lowered. (a) In the axial state, azimuthal modes with the same $\abs{\nu}$ are already split by the topological Aharonov-Bohm effect, which is unaffected by lowering the rotational symmetry. The spatial mode profiles show that the azimuthal modes remain running waves. (b) In the vortex-state, the singlets resulting from the doublet split due to symmetry lowering are categorized and colored according to their irreducible representation (irrep). Lines are added to guide the eye. Fully symmetric $A_1$ and fully antisymmetric $A_2$ singlets (blue) appear for $\abs{\nu}=c,2c,...$. Mixed-symmetry $B_1$ and $B_2$ singlets (orange) appear for $\abs{\nu}=c/2,3c/2,...$, but only in tubes with an even number of facets. Panels (c-d) show examples of the corresponding numerically-obtained lateral mode profiles (magnitude and phase of the $z$ component) as well as representative (anti-)mirror planes, distinguishing modes belonging to different irreps. The energetic split between two singlets with the same $\abs{\nu}$ can be understood from the mode localization following from these mirror symmetries (see main text).}
    \label{fig:FIG3}
\end{figure*}

Before confirming these symmetry-based predictions by micromagnetic simulations, we want to disentangle symmetry effects from topological effects. Namely, we show that the previously mentioned Aharonov-Bohm doublet splitting for magnons, which lifts the degeneracy between clockwise and counter-clockwise modes ($\mp \nu$), remains unaffected when lowering the rotational symmetry. Let us only consider the spin waves at $k=0$, which are homogeneous along the tubes. As known for round magnetic nanotubes ($c=\infty$) \cite{salazar-cardonaNonreciprocitySpinWaves2021}, the topological doublet splitting for modes with opposite azimuthal periods ($\nu$) is maximum when the tubes are axially magnetized. Here, the tubes are saturated in this direction using an external field of \SI{200}{\milli\tesla}. Fig.~\figref{fig:FIG3}{a} shows the frequencies of the modes at $k=0$, as the number of facets $c$ is decreased. In this case, the deformation of the cylindrical tube to a polygonal shape, does not have a major impact due to the topological origin of the Aharonov-Bohm splitting. Clearly, the topology of polygonal tubes is exactly the same as of round tubes.
Note, that the singlets of the topological split are still lateral-running waves, just like in round nanotubes. Thus, the phase of the modes varies smoothly along the circumference, as clear from the profiles of the modes with $\nu=-1$ periods in Fig.~\figref{fig:FIG3}{a}. 


In the vortex state, the Aharonov-Bohm flux is zero, therefore, in round tubes, clockwise and counter-clockwise ($\pm \nu$) modes are degenerate~\cite{otaloraCurvatureInducedAsymmetricSpinWave2016}. However, this degeneracy is lifted for some of the modes, when the rotational symmetry is lowered. Fig.~\figref{fig:FIG3}{b} shows the computed modes at $k=0$ for different numbers of facets. For certain modes, namely when the azimuthal periods are half-integer multiples ($\abs{\nu}=c/2$, $3c/2$, ...) or integer multiples ($\abs{\nu}=c$, $2c$, ...) of the number of facets, a splitting of the doublets is observed. This geometrical splitting is solely induced by lowering the rotational symmetry of the vortex state.

When classifying the modes according to the irreps (see Table I, Suppl. material S2), the singlets resulting from the symmetry-lowering-induced splitting appear in $A_1$--$A_2$ and $B_1$--$B_2$ pairs, colored in blue and orange in Fig.~\figref{fig:FIG3}{b}. This classification, using the symmetry of the magnitude and phase maps obtained from the dynamic-matrix approach~\cite{korberFiniteelementDynamicmatrixApproach2021} has the following simple meaning. 
The singlets with $\abs{\nu}=c,2c,...$ are either symmetric ($A_1$) or anti-symmetric ($A_2$) with respect to all mirror-plane reflections $m$, as seen in \figref{fig:FIG3}{c} for $\abs{\nu}=c$. The mirror and "anti-mirror" planes are indicated by solid and dashed lines, respectively. Naturally, the homogeneous mode ($\nu=0$) is a fully symmetric ($A_1$) singlet.
For tubes with even number of facets, two sets of mirror planes exist, corner-to-corner and facet-to-facet types. In this case, singlets symmetric to the former set and anti-symmetric to the latter set are called $B_1$. For $B_2$ the situation is reversed, as shown in \figref{fig:FIG3}{d} for $\abs{\nu}=c/2$.
Modes with such symmetry cannot exist for odd number of facets.

The modes in doublets with periods $\abs{\nu}\neq c/2,c,3c/2,2c,...$ are interchanged by some of the mirror symmetries. Therefore, they remain degenerate clockwise and counter-clockwise running modes, belonging to the two-dimensional irreps. In contrast, the singlets are lateral-standing waves, with a piecewise constant phase along the perimeter and with jumps by $\pi$ at certain points, see Fig.~\figref{fig:FIG3}{c,d}. 

In case of the doublet splittings observed here, the energy hierarchy of the resulting singlet shows a general trend: $A_1$ ($B_1$) singlets always have lower energy than the corresponding $A_2$ ($B_2$) singlets. This can be understood in a simple physical picture: Modes belonging to $A_2$ and $B_2$ representations have anti-mirror planes connecting opposite corners, where these modes must have a $\pi$ phase jump. As a result, the oscillation magnitude has to vanish at the corners and the modes are therefore localized to the facets. The opposite holds for modes belonging to the $A_1$ and $B_1$ representations. They possess mirror-planes connecting opposite corners, hence favor mode localization to the corners. Whether mode localization to corners or facets is energetically favored is not determined by the symmetry but the interactions in the system. In the case of polygonal vortices, the domains walls in the corners surrounded by the homogeneous domains in the flat facets [Fig.~\figref{fig:FIG2}{a}] act as potential wells for spin waves \cite{wagnerMagneticDomainWalls2016,korberSpinWaveReciprocityPresence2017}.

The doublet split induced by symmetry lowering is not restricted to the modes at $k=0$ but, occurs also for modes propagating along the tube axis. This is seen for selected polygonal tubes in the vortex state starting from the round tube ($c=\infty$) down to a hexagonal tube ($c=6$) in Fig.~\figref{fig:FIG4}{a-d}. 
Besides the dipole-induced dispersion asymmetry \cite{otaloraCurvatureInducedAsymmetricSpinWave2016,otaloraAsymmetricSpinwaveDispersion2017}, present for each $c$, the splitting due to symmetry lowering \textit{propagates} through the dispersion branches. Moreover, in the decagonal tube ($c=10$) an avoided level crossing close to $k=\SI{5}{\radian/\micro\meter}$ is present, which indicates the hybridization of two singlet branches. In the hexagonal tube ($c=6$) there are even three hybridized branches observable. In ferromagnetic samples, spin waves with different spatial profiles can be hybridized due to the presence of dynamic dipolar fields. This hybridization is mediated by the dipolar magnetostatic potential generated by each individual mode and its strength is determined by the spatial overlap of the \textit{unhybridized} modes, \textit{i.e.} the eigenmodes without dipolar interaction. In general, disentangling hybridized spin-wave modes for arbitrary geometries can be achieved numerically, as reported recently \cite{korberNumericalReverseEngineering2021a}. 
To tell if two crossing branches can hybridize or not it is enough to consider the symmetry of the modes.

\begin{figure}[h!]
    \centering
    \includegraphics{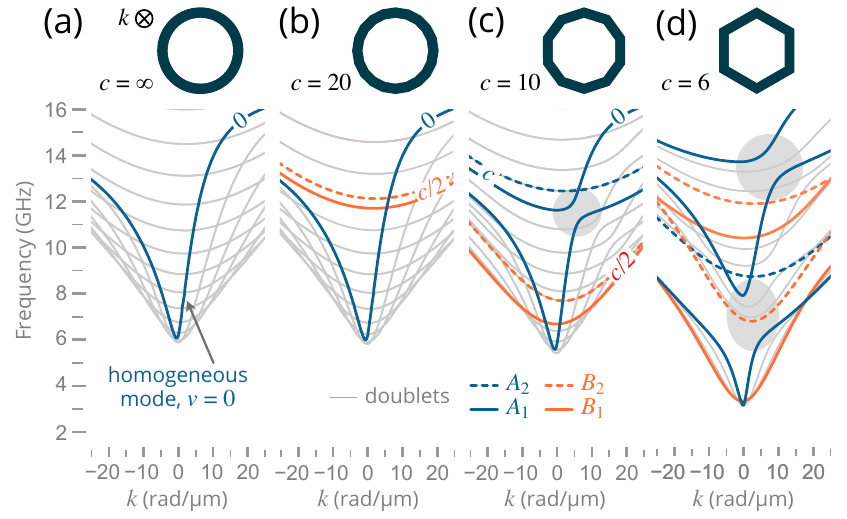}
    \caption{Spin-wave dispersions for different polygonal nanotubes in the vortex state, calculated using a propagating-wave dynamic matrix approach. All dispersions exhibit a curvature-induced asymmetry upon inversion of the wave vector $k$. Singlet branches are highlighted as thick colored (blue or orange) lines. As the number of corners $c$ decreases, the doublet split induced by symmetry lowering \textit{propagates} down in frequency and the singlets grow in number. Branches belonging to the same irreducible representation (irrep) make avoided level crossings due to dipole-dipole hybridization, highlighted by gray patches in (c) and (d).}
    \label{fig:FIG4}
\end{figure}

For a perfectly round tube, modes of different azimuthal periods $\nu$ and $\nu^\prime$, but even with the same wave vector $k$, belong to different irreps. Therefore, they are pairwise orthogonal, such that their volume-averaged inner product $\left\langle \bm{\eta}_{\nu,k}\cdot \bm{\eta}_{\nu^\prime,k} \right\rangle = 0$ vanishes. As a result, dipolar hybridization of different azimuthal branches is strictly forbidden. In vortex-state polygonal tubes with a finite number of corners, this orthogonality is broken. We have seen above that modes with azimuthal period being an integer multiple of the number of corners, $\abs{\nu}=nc$ (with $n = 1,2,...$), are arranged into either fully antisymmetric ($A_2$) or fully symmetric ($A_1$) lateral standing waves. The latter share the same symmetry as the fully homogeneous mode with $\nu=0$. Due to their common symmetry character, they can hybridize with each other in the presence of dynamic dipolar fields. This explains the avoided level crossings in the dispersion of the decagonal and the hexagonal tube in Fig.~\figref{fig:FIG4}{c,d}, where the homogeneous mode $\nu=0$ is hybridized with the lower-frequency $A_1$ modes localized to the corners, of the $\abs{\nu}=c$ and $\abs{\nu}=2c$ pairs (drawn as solid blue lines). More generally, modes belonging to the same irreps can be hybridized via dipolar interaction. An animated movie in the Suppl. material shows the transition of the dispersion for all numbers of facets between 30 and 6. This movie provides a clear visualization of how the avoided level crossings \textit{follow} the branches, that are split by lowering the rotational symmetry, i.e. they follow the singlet lines seen in Fig.~\figref{fig:FIG3}{b}. 


Similar arguments apply when discussing the susceptibility of the spin waves within polygonal tubes with respect to the resonant excitation by high-frequency external fields. To selectively excite certain branches of the dispersion using a microwave antenna, knowledge about the symmetry of the modes is paramount. In fact, when using an antenna which is not adapted to the symmetry of the waveguide, the excitation of undesired modes can hardly be avoided \cite{korberSymmetryCurvatureEffects2021}. Here, we briefly discuss the case of a current-loop antenna which possesses the same symmetry as the magnetic system itself, therefore belonging to the $A_1$ irrep [see Fig.~\figref{fig:FIG5}{a}]. In a perfectly round nanotube ($c=\infty$), such an antenna will only excite the homogeneous mode, $\nu=0$. However, in polygonal tubes there are additional $A_1$ modes, all of which should couple to such an antenna. This is seen in Fig.~\figref{fig:FIG5}{b}, for which we calculated the microwave absorption from the mode profiles according to Ref.~\citenum{korberSymmetryCurvatureEffects2021} as a function of $c$, taking into account the field distribution of the current-loop antennae wrapped around the different polygonal tubes.

\begin{figure}[h!]
    \centering
    \includegraphics{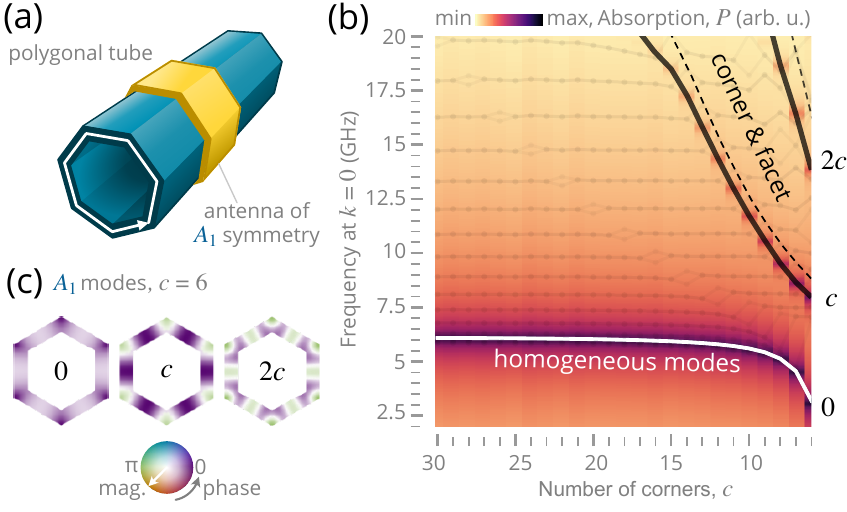}
    \caption{(a) Schematics of a current-loop antenna wrapped around a polygonal nanotube in the vortex state. (b) Corresponding microwave absorption at $k=0$ calculated from the mode profiles according to Eq. 2 in suppl. materials. Only modes with $A_1$ symmetry (lower-frequency singlets with $\abs{\nu}=0,c,2c,...$) can be excited with this antenna geometry. (c) Spatial profiles ($z$ component of magnitude and phase) of the corresponding $A_1$ modes for $c=6$.}
    \label{fig:FIG5}
\end{figure}

%
%

As seen in Fig.~\figref{fig:FIG5}{b}, next to the homogeneous mode, also the lower-frequency $A_1$ singlets with $\abs{\nu}=c, 2c, ...$ can be excited. The mode profiles of the $A_1$ modes for $c=6$ are seen in Fig.~\figref{fig:FIG5}{c}. When other branches of the dispersion are desired to be selectively excited, the microwave antenna has to be designed according to the symmetry of those modes.

%

In conclusion, we investigated the influence of discrete symmetries on the spin-wave dispersion in polygonal nanotubes by micromagnetic modeling. We found that lowering the rotational symmetry leads to the splitting of doublet modes. This is a purely geometrical effect and should be distinguished from the nontrivial version of the Aharonov-Bohm effect, more commonly known as the Berry phase of spin waves.
The splitting is observed only for certain azimuthal modes, when the azimuthal periods are half-integer or integer multiples of the number of facets. The resulting singlets are lateral-standing spin waves. Using symmetry analysis, the normal modes can be categorized according to the irreps of the magnetic point group defined by the geometry and the magnetic state of the tube. As a consequence of discrete symmetry, multiple modes can belong to the same irrep, allowing for the hybridization of modes with different azimuthal periods. This leads to a spin-wave dispersion with multiple avoided level crossings. Knowing the possible symmetries of the modes offers a qualitative understanding of the mode mixing. The knowledge about the symmetry of the modes is also important for the antenna design and applications, as we demonstrate by computing the absorption for a current loop antenna. Selective excitation of spin-wave branches is possible by designing the antenna according to the symmetry of the modes. We believe that these novel phenomena activated in discrete symmetries are important for a broad range of 2- and 3-dimensional nanostructures.


As a general conclusion of this study we can say that the dynamic properties of nanomaterials are determined by three general factors: 1) their geometrical shape, 2) the type of interactions governing the dynamical degrees of freedom and 3) the nature of their ground state. In the present study, as a specific example, we investigate polygonal tubes (1) in the ferromagnetic vortex state (3), where the spins interact via the Heisenberg exchange and the dipole-dipole interaction (2). We explore the importance of the structure--functionality relation in such nanotubes, clearly manifested in their magnetization dynamics via geometrically-induced mode splittings and symmetry-governed hybridization effects. We emphasize that this proof-of-concept study can be generalized to a wide range of nanomaterials. For example, the impact of discrete symmetries can be explored similarly, when spherical nanoparticles are replaced by polyhedral ones. Moreover, besides the dynamic magnetic properties studied here, one can similarly use geometrical effects to design elastic, dielectric and optical properties in nanomaterials, since acoustic and polarization waves are also prone to mode splitting and hybridization when lowering the symmetry of the host material.

Financial support by the Deutsche Forschungsgemeinschaft within the programs KA 5069/1-1 and KA 5069/3-1 is gratefully acknowledged.


\begin{thebibliography}{35}%
\makeatletter
\providecommand \@ifxundefined [1]{%
 \@ifx{#1\undefined}
}%
\providecommand \@ifnum [1]{%
 \ifnum #1\expandafter \@firstoftwo
 \else \expandafter \@secondoftwo
 \fi
}%
\providecommand \@ifx [1]{%
 \ifx #1\expandafter \@firstoftwo
 \else \expandafter \@secondoftwo
 \fi
}%
\providecommand \natexlab [1]{#1}%
\providecommand \enquote  [1]{``#1''}%
\providecommand \bibnamefont  [1]{#1}%
\providecommand \bibfnamefont [1]{#1}%
\providecommand \citenamefont [1]{#1}%
\providecommand \href@noop [0]{\@secondoftwo}%
\providecommand \href [0]{\begingroup \@sanitize@url \@href}%
\providecommand \@href[1]{\@@startlink{#1}\@@href}%
\providecommand \@@href[1]{\endgroup#1\@@endlink}%
\providecommand \@sanitize@url [0]{\catcode `\\12\catcode `\$12\catcode
  `\&12\catcode `\#12\catcode `\^12\catcode `\_12\catcode `\%12\relax}%
\providecommand \@@startlink[1]{}%
\providecommand \@@endlink[0]{}%
\providecommand \url  [0]{\begingroup\@sanitize@url \@url }%
\providecommand \@url [1]{\endgroup\@href {#1}{\urlprefix }}%
\providecommand \urlprefix  [0]{URL }%
\providecommand \Eprint [0]{\href }%
\providecommand \doibase [0]{https://doi.org/}%
\providecommand \selectlanguage [0]{\@gobble}%
\providecommand \bibinfo  [0]{\@secondoftwo}%
\providecommand \bibfield  [0]{\@secondoftwo}%
\providecommand \translation [1]{[#1]}%
\providecommand \BibitemOpen [0]{}%
\providecommand \bibitemStop [0]{}%
\providecommand \bibitemNoStop [0]{.\EOS\space}%
\providecommand \EOS [0]{\spacefactor3000\relax}%
\providecommand \BibitemShut  [1]{\csname bibitem#1\endcsname}%
\let\auto@bib@innerbib\@empty
\bibitem [{\citenamefont {Smith}\ \emph {et~al.}(2011)\citenamefont {Smith},
  \citenamefont {Makarov}, \citenamefont {Sanchez}, \citenamefont {Fomin},\
  and\ \citenamefont {Schmidt}}]{Smith11}%
  \BibitemOpen
  \bibfield  {author} {\bibinfo {author} {\bibfnamefont {E.~J.}\ \bibnamefont
  {Smith}}, \bibinfo {author} {\bibfnamefont {D.}~\bibnamefont {Makarov}},
  \bibinfo {author} {\bibfnamefont {S.}~\bibnamefont {Sanchez}}, \bibinfo
  {author} {\bibfnamefont {V.~M.}\ \bibnamefont {Fomin}},\ and\ \bibinfo
  {author} {\bibfnamefont {O.~G.}\ \bibnamefont {Schmidt}},\ }\bibfield
  {title} {\bibinfo {title} {Magnetic microhelix coil structures},\ }\href
  {https://doi.org/10.1103/PhysRevLett.107.097204} {\bibfield  {journal}
  {\bibinfo  {journal} {Physical Review Letters}\ }\textbf {\bibinfo {volume}
  {107}},\ \bibinfo {pages} {097204} (\bibinfo {year} {2011})}\BibitemShut
  {NoStop}%
\bibitem [{\citenamefont {Volkov}\ \emph
  {et~al.}(2019{\natexlab{a}})\citenamefont {Volkov}, \citenamefont {K\'akay},
  \citenamefont {Kronast}, \citenamefont {M\"onch}, \citenamefont {Mawass},
  \citenamefont {Fassbender},\ and\ \citenamefont {Makarov}}]{Volkov19c}%
  \BibitemOpen
  \bibfield  {author} {\bibinfo {author} {\bibfnamefont {O.~M.}\ \bibnamefont
  {Volkov}}, \bibinfo {author} {\bibfnamefont {A.}~\bibnamefont {K\'akay}},
  \bibinfo {author} {\bibfnamefont {F.}~\bibnamefont {Kronast}}, \bibinfo
  {author} {\bibfnamefont {I.}~\bibnamefont {M\"onch}}, \bibinfo {author}
  {\bibfnamefont {M.-A.}\ \bibnamefont {Mawass}}, \bibinfo {author}
  {\bibfnamefont {J.}~\bibnamefont {Fassbender}},\ and\ \bibinfo {author}
  {\bibfnamefont {D.}~\bibnamefont {Makarov}},\ }\bibfield  {title} {\bibinfo
  {title} {Experimental observation of exchange-driven chiral effects in
  curvilinear magnetism},\ }\href
  {https://doi.org/10.1103/PhysRevLett.123.077201} {\bibfield  {journal}
  {\bibinfo  {journal} {Physical Review Letters}\ }\textbf {\bibinfo {volume}
  {123}},\ \bibinfo {pages} {077201} (\bibinfo {year}
  {2019}{\natexlab{a}})}\BibitemShut {NoStop}%
\bibitem [{\citenamefont {Volkov}\ \emph
  {et~al.}(2019{\natexlab{b}})\citenamefont {Volkov}, \citenamefont {Kronast},
  \citenamefont {M{\"{o}}nch}, \citenamefont {Mawass}, \citenamefont
  {K{\'{a}}kay}, \citenamefont {Fassbender},\ and\ \citenamefont
  {Makarov}}]{Volkov19}%
  \BibitemOpen
  \bibfield  {author} {\bibinfo {author} {\bibfnamefont {O.~M.}\ \bibnamefont
  {Volkov}}, \bibinfo {author} {\bibfnamefont {F.}~\bibnamefont {Kronast}},
  \bibinfo {author} {\bibfnamefont {I.}~\bibnamefont {M{\"{o}}nch}}, \bibinfo
  {author} {\bibfnamefont {M.-A.}\ \bibnamefont {Mawass}}, \bibinfo {author}
  {\bibfnamefont {A.}~\bibnamefont {K{\'{a}}kay}}, \bibinfo {author}
  {\bibfnamefont {J.}~\bibnamefont {Fassbender}},\ and\ \bibinfo {author}
  {\bibfnamefont {D.}~\bibnamefont {Makarov}},\ }\bibfield  {title} {\bibinfo
  {title} {Experimental and theoretical study of curvature effects in parabolic
  nanostripes},\ }\href {https://doi.org/10.1002/pssr.201800309} {\bibfield
  {journal} {\bibinfo  {journal} {Physica Status Solidi ({RRL}) - Rapid
  Research Letters}\ }\textbf {\bibinfo {volume} {13}},\ \bibinfo {pages}
  {1800309} (\bibinfo {year} {2019}{\natexlab{b}})}\BibitemShut {NoStop}%
\bibitem [{\citenamefont {Streubel}\ \emph {et~al.}(2012)\citenamefont
  {Streubel}, \citenamefont {Kravchuk}, \citenamefont {Sheka}, \citenamefont
  {Makarov}, \citenamefont {Kronast}, \citenamefont {Schmidt},\ and\
  \citenamefont {Gaididei}}]{Streubel12}%
  \BibitemOpen
  \bibfield  {author} {\bibinfo {author} {\bibfnamefont {R.}~\bibnamefont
  {Streubel}}, \bibinfo {author} {\bibfnamefont {V.~P.}\ \bibnamefont
  {Kravchuk}}, \bibinfo {author} {\bibfnamefont {D.~D.}\ \bibnamefont {Sheka}},
  \bibinfo {author} {\bibfnamefont {D.}~\bibnamefont {Makarov}}, \bibinfo
  {author} {\bibfnamefont {F.}~\bibnamefont {Kronast}}, \bibinfo {author}
  {\bibfnamefont {O.~G.}\ \bibnamefont {Schmidt}},\ and\ \bibinfo {author}
  {\bibfnamefont {Y.}~\bibnamefont {Gaididei}},\ }\bibfield  {title} {\bibinfo
  {title} {Equilibrium magnetic states in individual hemispherical permalloy
  caps},\ }\href {https://doi.org/10.1063/1.4756708} {\bibfield  {journal}
  {\bibinfo  {journal} {Applied Physics Letters}\ }\textbf {\bibinfo {volume}
  {101}},\ \bibinfo {eid} {132419} (\bibinfo {year} {2012})}\BibitemShut
  {NoStop}%
\bibitem [{\citenamefont {Streubel}\ \emph {et~al.}(2013)\citenamefont
  {Streubel}, \citenamefont {Makarov}, \citenamefont {Lee}, \citenamefont
  {M{\"u}ller}, \citenamefont {Melzer}, \citenamefont {Sch{\"a}fer},
  \citenamefont {Bufon}, \citenamefont {Kim},\ and\ \citenamefont
  {Schmidt}}]{Streubel13c}%
  \BibitemOpen
  \bibfield  {author} {\bibinfo {author} {\bibfnamefont {R.}~\bibnamefont
  {Streubel}}, \bibinfo {author} {\bibfnamefont {D.}~\bibnamefont {Makarov}},
  \bibinfo {author} {\bibfnamefont {J.}~\bibnamefont {Lee}}, \bibinfo {author}
  {\bibfnamefont {C.}~\bibnamefont {M{\"u}ller}}, \bibinfo {author}
  {\bibfnamefont {M.}~\bibnamefont {Melzer}}, \bibinfo {author} {\bibfnamefont
  {R.}~\bibnamefont {Sch{\"a}fer}}, \bibinfo {author} {\bibfnamefont
  {C.~C.~B.}\ \bibnamefont {Bufon}}, \bibinfo {author} {\bibfnamefont {S.-K.}\
  \bibnamefont {Kim}},\ and\ \bibinfo {author} {\bibfnamefont {O.~G.}\
  \bibnamefont {Schmidt}},\ }\bibfield  {title} {\bibinfo {title} {Rolled-up
  permalloy nanomembranes with multiple windings},\ }\href
  {https://doi.org/10.1142/s2010324713400018} {\bibfield  {journal} {\bibinfo
  {journal} {SPIN}\ }\textbf {\bibinfo {volume} {03}},\ \bibinfo {pages}
  {1340001} (\bibinfo {year} {2013})}\BibitemShut {NoStop}%
\bibitem [{\citenamefont {Kravchuk}\ \emph {et~al.}(2016)\citenamefont
  {Kravchuk}, \citenamefont {R\"o\ss{}ler}, \citenamefont {Volkov},
  \citenamefont {Sheka}, \citenamefont {van~den Brink}, \citenamefont
  {Makarov}, \citenamefont {Fuchs}, \citenamefont {Fangohr},\ and\
  \citenamefont {Gaididei}}]{KravchukPRB2016SphericalShell}%
  \BibitemOpen
  \bibfield  {author} {\bibinfo {author} {\bibfnamefont {V.~P.}\ \bibnamefont
  {Kravchuk}}, \bibinfo {author} {\bibfnamefont {U.~K.}\ \bibnamefont
  {R\"o\ss{}ler}}, \bibinfo {author} {\bibfnamefont {O.~M.}\ \bibnamefont
  {Volkov}}, \bibinfo {author} {\bibfnamefont {D.~D.}\ \bibnamefont {Sheka}},
  \bibinfo {author} {\bibfnamefont {J.}~\bibnamefont {van~den Brink}}, \bibinfo
  {author} {\bibfnamefont {D.}~\bibnamefont {Makarov}}, \bibinfo {author}
  {\bibfnamefont {H.}~\bibnamefont {Fuchs}}, \bibinfo {author} {\bibfnamefont
  {H.}~\bibnamefont {Fangohr}},\ and\ \bibinfo {author} {\bibfnamefont
  {Y.}~\bibnamefont {Gaididei}},\ }\bibfield  {title} {\bibinfo {title}
  {Topologically stable magnetization states on a spherical shell:
  Curvature-stabilized skyrmions},\ }\href
  {https://doi.org/10.1103/PhysRevB.94.144402} {\bibfield  {journal} {\bibinfo
  {journal} {Phys. Rev. B}\ }\textbf {\bibinfo {volume} {94}},\ \bibinfo
  {pages} {144402} (\bibinfo {year} {2016})}\BibitemShut {NoStop}%
\bibitem [{\citenamefont {Elías}\ \emph {et~al.}(2019)\citenamefont {Elías},
  \citenamefont {Vidal-Silva},\ and\ \citenamefont
  {Carvalho-Santos}}]{elias_winding_2019}%
  \BibitemOpen
  \bibfield  {author} {\bibinfo {author} {\bibfnamefont {R.~G.}\ \bibnamefont
  {Elías}}, \bibinfo {author} {\bibfnamefont {N.}~\bibnamefont
  {Vidal-Silva}},\ and\ \bibinfo {author} {\bibfnamefont {V.~L.}\ \bibnamefont
  {Carvalho-Santos}},\ }\bibfield  {title} {\bibinfo {title} {Winding number
  selection on merons by {Gaussian} curvature’s sign},\ }\href
  {https://doi.org/10.1038/s41598-019-50395-7} {\bibfield  {journal} {\bibinfo
  {journal} {Scientific Reports}\ }\textbf {\bibinfo {volume} {9}},\ \bibinfo
  {pages} {14309} (\bibinfo {year} {2019})}\BibitemShut {NoStop}%
\bibitem [{\citenamefont {Yan}\ \emph {et~al.}(2011)\citenamefont {Yan},
  \citenamefont {Andreas}, \citenamefont {K{\'a}kay}, \citenamefont
  {Garc{\'i}a-S{\'a}nchez},\ and\ \citenamefont {Hertel}}]{Yan11a}%
  \BibitemOpen
  \bibfield  {author} {\bibinfo {author} {\bibfnamefont {M.}~\bibnamefont
  {Yan}}, \bibinfo {author} {\bibfnamefont {C.}~\bibnamefont {Andreas}},
  \bibinfo {author} {\bibfnamefont {A.}~\bibnamefont {K{\'a}kay}}, \bibinfo
  {author} {\bibfnamefont {F.}~\bibnamefont {Garc{\'i}a-S{\'a}nchez}},\ and\
  \bibinfo {author} {\bibfnamefont {R.}~\bibnamefont {Hertel}},\ }\bibfield
  {title} {\bibinfo {title} {{Fast domain wall dynamics in magnetic nanotubes:
  Suppression of Walker breakdown and Cherenkov-like spin wave emission}},\
  }\href {https://doi.org/10.1063/1.3643037} {\bibfield  {journal} {\bibinfo
  {journal} {Applied Physics Letters}\ }\textbf {\bibinfo {volume} {99}},\
  \bibinfo {eid} {122505} (\bibinfo {year} {2011})}\BibitemShut {NoStop}%
\bibitem [{\citenamefont {Yan}\ \emph {et~al.}(2013)\citenamefont {Yan},
  \citenamefont {K\'akay}, \citenamefont {Andreas},\ and\ \citenamefont
  {Hertel}}]{Yan13}%
  \BibitemOpen
  \bibfield  {author} {\bibinfo {author} {\bibfnamefont {M.}~\bibnamefont
  {Yan}}, \bibinfo {author} {\bibfnamefont {A.}~\bibnamefont {K\'akay}},
  \bibinfo {author} {\bibfnamefont {C.}~\bibnamefont {Andreas}},\ and\ \bibinfo
  {author} {\bibfnamefont {R.}~\bibnamefont {Hertel}},\ }\bibfield  {title}
  {\bibinfo {title} {{Spin-Cherenkov effect and magnonic Mach cones}},\ }\href
  {https://doi.org/10.1103/PhysRevB.88.220412} {\bibfield  {journal} {\bibinfo
  {journal} {Physical Review B}\ }\textbf {\bibinfo {volume} {88}},\ \bibinfo
  {pages} {220412} (\bibinfo {year} {2013})}\BibitemShut {NoStop}%
\bibitem [{\citenamefont {Ot{\'a}lora}\ \emph {et~al.}(2016)\citenamefont
  {Ot{\'a}lora}, \citenamefont {Yan}, \citenamefont {Schultheiss},
  \citenamefont {Hertel},\ and\ \citenamefont
  {K{\'a}kay}}]{otaloraCurvatureInducedAsymmetricSpinWave2016}%
  \BibitemOpen
  \bibfield  {author} {\bibinfo {author} {\bibfnamefont {J.~A.}\ \bibnamefont
  {Ot{\'a}lora}}, \bibinfo {author} {\bibfnamefont {M.}~\bibnamefont {Yan}},
  \bibinfo {author} {\bibfnamefont {H.}~\bibnamefont {Schultheiss}}, \bibinfo
  {author} {\bibfnamefont {R.}~\bibnamefont {Hertel}},\ and\ \bibinfo {author}
  {\bibfnamefont {A.}~\bibnamefont {K{\'a}kay}},\ }\bibfield  {title} {\bibinfo
  {title} {Curvature-{{Induced Asymmetric Spin}}-{{Wave Dispersion}}},\ }\href
  {https://doi.org/10.1103/PhysRevLett.117.227203} {\bibfield  {journal}
  {\bibinfo  {journal} {Physical Review Letters}\ }\textbf {\bibinfo {volume}
  {117}},\ \bibinfo {pages} {227203} (\bibinfo {year} {2016})}\BibitemShut
  {NoStop}%
\bibitem [{\citenamefont {Bordács}\ \emph {et~al.}(2012)\citenamefont
  {Bordács}, \citenamefont {Kézsmárki}, \citenamefont {Szaller},
  \citenamefont {Demkó}, \citenamefont {Kida}, \citenamefont {Murakawa},
  \citenamefont {Onose}, \citenamefont {Shimano}, \citenamefont {Rõõm},
  \citenamefont {Nagel}, \citenamefont {Miyahara}, \citenamefont {Furukawa},\
  and\ \citenamefont {Tokura}}]{bordacs_chirality_2012}%
  \BibitemOpen
  \bibfield  {author} {\bibinfo {author} {\bibfnamefont {S.}~\bibnamefont
  {Bordács}}, \bibinfo {author} {\bibfnamefont {I.}~\bibnamefont
  {Kézsmárki}}, \bibinfo {author} {\bibfnamefont {D.}~\bibnamefont
  {Szaller}}, \bibinfo {author} {\bibfnamefont {L.}~\bibnamefont {Demkó}},
  \bibinfo {author} {\bibfnamefont {N.}~\bibnamefont {Kida}}, \bibinfo {author}
  {\bibfnamefont {H.}~\bibnamefont {Murakawa}}, \bibinfo {author}
  {\bibfnamefont {Y.}~\bibnamefont {Onose}}, \bibinfo {author} {\bibfnamefont
  {R.}~\bibnamefont {Shimano}}, \bibinfo {author} {\bibfnamefont
  {T.}~\bibnamefont {Rõõm}}, \bibinfo {author} {\bibfnamefont
  {U.}~\bibnamefont {Nagel}}, \bibinfo {author} {\bibfnamefont
  {S.}~\bibnamefont {Miyahara}}, \bibinfo {author} {\bibfnamefont
  {N.}~\bibnamefont {Furukawa}},\ and\ \bibinfo {author} {\bibfnamefont
  {Y.}~\bibnamefont {Tokura}},\ }\bibfield  {title} {\bibinfo {title}
  {Chirality of matter shows up via spin excitations},\ }\href
  {https://doi.org/10.1038/nphys2387} {\bibfield  {journal} {\bibinfo
  {journal} {Nature Physics}\ }\textbf {\bibinfo {volume} {8}},\ \bibinfo
  {pages} {734} (\bibinfo {year} {2012})}\BibitemShut {NoStop}%
\bibitem [{\citenamefont {Kézsmárki}\ \emph {et~al.}(2014)\citenamefont
  {Kézsmárki}, \citenamefont {Szaller}, \citenamefont {Bordács},
  \citenamefont {Kocsis}, \citenamefont {Tokunaga}, \citenamefont {Taguchi},
  \citenamefont {Murakawa}, \citenamefont {Tokura}, \citenamefont {Engelkamp},
  \citenamefont {Rõõm},\ and\ \citenamefont
  {Nagel}}]{kezsmarki_one-way_2014}%
  \BibitemOpen
  \bibfield  {author} {\bibinfo {author} {\bibfnamefont {I.}~\bibnamefont
  {Kézsmárki}}, \bibinfo {author} {\bibfnamefont {D.}~\bibnamefont
  {Szaller}}, \bibinfo {author} {\bibfnamefont {S.}~\bibnamefont {Bordács}},
  \bibinfo {author} {\bibfnamefont {V.}~\bibnamefont {Kocsis}}, \bibinfo
  {author} {\bibfnamefont {Y.}~\bibnamefont {Tokunaga}}, \bibinfo {author}
  {\bibfnamefont {Y.}~\bibnamefont {Taguchi}}, \bibinfo {author} {\bibfnamefont
  {H.}~\bibnamefont {Murakawa}}, \bibinfo {author} {\bibfnamefont
  {Y.}~\bibnamefont {Tokura}}, \bibinfo {author} {\bibfnamefont
  {H.}~\bibnamefont {Engelkamp}}, \bibinfo {author} {\bibfnamefont
  {T.}~\bibnamefont {Rõõm}},\ and\ \bibinfo {author} {\bibfnamefont
  {U.}~\bibnamefont {Nagel}},\ }\bibfield  {title} {\bibinfo {title} {One-way
  transparency of four-coloured spin-wave excitations in multiferroic
  materials},\ }\href {https://doi.org/10.1038/ncomms4203} {\bibfield
  {journal} {\bibinfo  {journal} {Nature Communications}\ }\textbf {\bibinfo
  {volume} {5}},\ \bibinfo {pages} {3203} (\bibinfo {year} {2014})}\BibitemShut
  {NoStop}%
\bibitem [{\citenamefont {Gaididei}\ \emph {et~al.}(2014)\citenamefont
  {Gaididei}, \citenamefont {Kravchuk},\ and\ \citenamefont
  {Sheka}}]{GaidideiPRL_curvature_effects_2014}%
  \BibitemOpen
  \bibfield  {author} {\bibinfo {author} {\bibfnamefont {Y.}~\bibnamefont
  {Gaididei}}, \bibinfo {author} {\bibfnamefont {V.~P.}\ \bibnamefont
  {Kravchuk}},\ and\ \bibinfo {author} {\bibfnamefont {D.~D.}\ \bibnamefont
  {Sheka}},\ }\bibfield  {title} {\bibinfo {title} {Curvature effects in thin
  magnetic shells},\ }\href {https://doi.org/10.1103/PhysRevLett.112.257203}
  {\bibfield  {journal} {\bibinfo  {journal} {Phys. Rev. Lett.}\ }\textbf
  {\bibinfo {volume} {112}},\ \bibinfo {pages} {257203} (\bibinfo {year}
  {2014})}\BibitemShut {NoStop}%
\bibitem [{\citenamefont {Sheka}\ \emph {et~al.}(2015)\citenamefont {Sheka},
  \citenamefont {Kravchuk},\ and\ \citenamefont {Gaididei}}]{Sheka_2015}%
  \BibitemOpen
  \bibfield  {author} {\bibinfo {author} {\bibfnamefont {D.~D.}\ \bibnamefont
  {Sheka}}, \bibinfo {author} {\bibfnamefont {V.~P.}\ \bibnamefont
  {Kravchuk}},\ and\ \bibinfo {author} {\bibfnamefont {Y.}~\bibnamefont
  {Gaididei}},\ }\bibfield  {title} {\bibinfo {title} {Curvature effects in
  statics and dynamics of low dimensional magnets},\ }\href
  {https://doi.org/10.1088/1751-8113/48/12/125202} {\bibfield  {journal}
  {\bibinfo  {journal} {Journal of Physics A: Mathematical and Theoretical}\
  }\textbf {\bibinfo {volume} {48}},\ \bibinfo {pages} {125202} (\bibinfo
  {year} {2015})}\BibitemShut {NoStop}%
\bibitem [{\citenamefont {Hill}\ \emph {et~al.}(2021)\citenamefont {Hill},
  \citenamefont {Slastikov},\ and\ \citenamefont
  {Tchernyshyov}}]{10.21468/SciPostPhys.10.3.078}%
  \BibitemOpen
  \bibfield  {author} {\bibinfo {author} {\bibfnamefont {D.}~\bibnamefont
  {Hill}}, \bibinfo {author} {\bibfnamefont {V.}~\bibnamefont {Slastikov}},\
  and\ \bibinfo {author} {\bibfnamefont {O.}~\bibnamefont {Tchernyshyov}},\
  }\bibfield  {title} {\bibinfo {title} {{Chiral magnetism: a geometric
  perspective}},\ }\href {https://doi.org/10.21468/SciPostPhys.10.3.078}
  {\bibfield  {journal} {\bibinfo  {journal} {SciPost Phys.}\ }\textbf
  {\bibinfo {volume} {10}},\ \bibinfo {pages} {78} (\bibinfo {year}
  {2021})}\BibitemShut {NoStop}%
\bibitem [{\citenamefont {Sheka}\ \emph {et~al.}(2020)\citenamefont {Sheka},
  \citenamefont {Pylypovskyi}, \citenamefont {Landeros}, \citenamefont
  {Gaididei}, \citenamefont {Kákay},\ and\ \citenamefont
  {Makarov}}]{sheka_nonlocal_2020}%
  \BibitemOpen
  \bibfield  {author} {\bibinfo {author} {\bibfnamefont {D.~D.}\ \bibnamefont
  {Sheka}}, \bibinfo {author} {\bibfnamefont {O.~V.}\ \bibnamefont
  {Pylypovskyi}}, \bibinfo {author} {\bibfnamefont {P.}~\bibnamefont
  {Landeros}}, \bibinfo {author} {\bibfnamefont {Y.}~\bibnamefont {Gaididei}},
  \bibinfo {author} {\bibfnamefont {A.}~\bibnamefont {Kákay}},\ and\ \bibinfo
  {author} {\bibfnamefont {D.}~\bibnamefont {Makarov}},\ }\bibfield  {title}
  {{\selectlanguage {en}\bibinfo {title} {Nonlocal chiral symmetry breaking in
  curvilinear magnetic shells}},\ }\href
  {https://doi.org/10.1038/s42005-020-0387-2} {\bibfield  {journal} {\bibinfo
  {journal} {Communications Physics}\ }\textbf {\bibinfo {volume} {3}},\
  \bibinfo {pages} {128} (\bibinfo {year} {2020})}\BibitemShut {NoStop}%
\bibitem [{\citenamefont {Szaller}\ \emph {et~al.}(2013)\citenamefont
  {Szaller}, \citenamefont {Bord\'acs},\ and\ \citenamefont
  {K\'ezsm\'arki}}]{Szaller2013_Symmetry}%
  \BibitemOpen
  \bibfield  {author} {\bibinfo {author} {\bibfnamefont {D.}~\bibnamefont
  {Szaller}}, \bibinfo {author} {\bibfnamefont {S.}~\bibnamefont {Bord\'acs}},\
  and\ \bibinfo {author} {\bibfnamefont {I.}~\bibnamefont {K\'ezsm\'arki}},\
  }\bibfield  {title} {\bibinfo {title} {Symmetry conditions for nonreciprocal
  light propagation in magnetic crystals},\ }\href
  {https://doi.org/10.1103/PhysRevB.87.014421} {\bibfield  {journal} {\bibinfo
  {journal} {Phys. Rev. B}\ }\textbf {\bibinfo {volume} {87}},\ \bibinfo
  {pages} {014421} (\bibinfo {year} {2013})}\BibitemShut {NoStop}%
\bibitem [{\citenamefont {Udvardi}\ and\ \citenamefont
  {Szunyogh}(2009)}]{Udvardi2009_chiral_asymmetry}%
  \BibitemOpen
  \bibfield  {author} {\bibinfo {author} {\bibfnamefont {L.}~\bibnamefont
  {Udvardi}}\ and\ \bibinfo {author} {\bibfnamefont {L.}~\bibnamefont
  {Szunyogh}},\ }\bibfield  {title} {\bibinfo {title} {Chiral asymmetry of the
  spin-wave spectra in ultrathin magnetic films},\ }\href
  {https://doi.org/10.1103/PhysRevLett.102.207204} {\bibfield  {journal}
  {\bibinfo  {journal} {Phys. Rev. Lett.}\ }\textbf {\bibinfo {volume} {102}},\
  \bibinfo {pages} {207204} (\bibinfo {year} {2009})}\BibitemShut {NoStop}%
\bibitem [{\citenamefont {Pylypovskyi}\ \emph {et~al.}(2015)\citenamefont
  {Pylypovskyi}, \citenamefont {Kravchuk}, \citenamefont {Sheka}, \citenamefont
  {Makarov}, \citenamefont {Schmidt},\ and\ \citenamefont
  {Gaididei}}]{Pylypovskyi15b}%
  \BibitemOpen
  \bibfield  {author} {\bibinfo {author} {\bibfnamefont {O.~V.}\ \bibnamefont
  {Pylypovskyi}}, \bibinfo {author} {\bibfnamefont {V.~P.}\ \bibnamefont
  {Kravchuk}}, \bibinfo {author} {\bibfnamefont {D.~D.}\ \bibnamefont {Sheka}},
  \bibinfo {author} {\bibfnamefont {D.}~\bibnamefont {Makarov}}, \bibinfo
  {author} {\bibfnamefont {O.~G.}\ \bibnamefont {Schmidt}},\ and\ \bibinfo
  {author} {\bibfnamefont {Y.}~\bibnamefont {Gaididei}},\ }\bibfield  {title}
  {\bibinfo {title} {Coupling of chiralities in spin and physical spaces: {T}he
  {M}\"obius ring as a case study},\ }\href
  {https://doi.org/10.1103/PhysRevLett.114.197204} {\bibfield  {journal}
  {\bibinfo  {journal} {Physical Review Letters}\ }\textbf {\bibinfo {volume}
  {114}},\ \bibinfo {pages} {197204} (\bibinfo {year} {2015})}\BibitemShut
  {NoStop}%
\bibitem [{\citenamefont {Dugaev}\ \emph {et~al.}(2005)\citenamefont {Dugaev},
  \citenamefont {Bruno}, \citenamefont {Canals},\ and\ \citenamefont
  {Lacroix}}]{dugaevBerryPhaseMagnons2005}%
  \BibitemOpen
  \bibfield  {author} {\bibinfo {author} {\bibfnamefont {V.~K.}\ \bibnamefont
  {Dugaev}}, \bibinfo {author} {\bibfnamefont {P.}~\bibnamefont {Bruno}},
  \bibinfo {author} {\bibfnamefont {B.}~\bibnamefont {Canals}},\ and\ \bibinfo
  {author} {\bibfnamefont {C.}~\bibnamefont {Lacroix}},\ }\bibfield  {title}
  {\bibinfo {title} {Berry phase of magnons in textured ferromagnets},\ }\href
  {https://doi.org/10.1103/PhysRevB.72.024456} {\bibfield  {journal} {\bibinfo
  {journal} {Physical Review B - Condensed Matter and Materials Physics}\
  }\textbf {\bibinfo {volume} {72}},\ \bibinfo {pages} {1} (\bibinfo {year}
  {2005})}\BibitemShut {NoStop}%
\bibitem [{\citenamefont {{Salazar-Cardona}}\ \emph {et~al.}(2021)\citenamefont
  {{Salazar-Cardona}}, \citenamefont {K{\"o}rber}, \citenamefont {Schultheiss},
  \citenamefont {Lenz}, \citenamefont {Thomas}, \citenamefont {Nielsch},
  \citenamefont {K{\'a}kay},\ and\ \citenamefont
  {Ot{\'a}lora}}]{salazar-cardonaNonreciprocitySpinWaves2021}%
  \BibitemOpen
  \bibfield  {author} {\bibinfo {author} {\bibfnamefont {M.~M.}\ \bibnamefont
  {{Salazar-Cardona}}}, \bibinfo {author} {\bibfnamefont {L.}~\bibnamefont
  {K{\"o}rber}}, \bibinfo {author} {\bibfnamefont {H.}~\bibnamefont
  {Schultheiss}}, \bibinfo {author} {\bibfnamefont {K.}~\bibnamefont {Lenz}},
  \bibinfo {author} {\bibfnamefont {A.}~\bibnamefont {Thomas}}, \bibinfo
  {author} {\bibfnamefont {K.}~\bibnamefont {Nielsch}}, \bibinfo {author}
  {\bibfnamefont {A.}~\bibnamefont {K{\'a}kay}},\ and\ \bibinfo {author}
  {\bibfnamefont {J.~A.}\ \bibnamefont {Ot{\'a}lora}},\ }\bibfield  {title}
  {\bibinfo {title} {Nonreciprocity of spin waves in magnetic nanotubes with
  helical equilibrium magnetization},\ }\href
  {https://doi.org/10.1063/5.0048692} {\bibfield  {journal} {\bibinfo
  {journal} {Applied Physics Letters}\ }\textbf {\bibinfo {volume} {118}},\
  \bibinfo {pages} {262411} (\bibinfo {year} {2021})}\BibitemShut {NoStop}%
\bibitem [{\citenamefont {Aharonov}\ and\ \citenamefont
  {Bohm}(1959)}]{AharonovBohm_PRL1959}%
  \BibitemOpen
  \bibfield  {author} {\bibinfo {author} {\bibfnamefont {Y.}~\bibnamefont
  {Aharonov}}\ and\ \bibinfo {author} {\bibfnamefont {D.}~\bibnamefont
  {Bohm}},\ }\bibfield  {title} {\bibinfo {title} {Significance of
  electromagnetic potentials in the quantum theory},\ }\href
  {https://doi.org/10.1103/PhysRev.115.485} {\bibfield  {journal} {\bibinfo
  {journal} {Phys. Rev.}\ }\textbf {\bibinfo {volume} {115}},\ \bibinfo {pages}
  {485} (\bibinfo {year} {1959})}\BibitemShut {NoStop}%
\bibitem [{\citenamefont {Ivanov}\ and\ \citenamefont
  {Zaspel}(2005)}]{Mode_split_Ivanon_2005}%
  \BibitemOpen
  \bibfield  {author} {\bibinfo {author} {\bibfnamefont {B.~A.}\ \bibnamefont
  {Ivanov}}\ and\ \bibinfo {author} {\bibfnamefont {C.~E.}\ \bibnamefont
  {Zaspel}},\ }\bibfield  {title} {\bibinfo {title} {High frequency modes in
  vortex-state nanomagnets},\ }\href
  {https://doi.org/10.1103/PhysRevLett.94.027205} {\bibfield  {journal}
  {\bibinfo  {journal} {Phys. Rev. Lett.}\ }\textbf {\bibinfo {volume} {94}},\
  \bibinfo {pages} {027205} (\bibinfo {year} {2005})}\BibitemShut {NoStop}%
\bibitem [{\citenamefont {Bachtold}\ \emph {et~al.}(1999)\citenamefont
  {Bachtold}, \citenamefont {Strunk}, \citenamefont {Salvetat}, \citenamefont
  {Bonard}, \citenamefont {Forró}, \citenamefont {Nussbaumer},\ and\
  \citenamefont {Schönenberger}}]{bachtold_aharonovbohm_1999}%
  \BibitemOpen
  \bibfield  {author} {\bibinfo {author} {\bibfnamefont {A.}~\bibnamefont
  {Bachtold}}, \bibinfo {author} {\bibfnamefont {C.}~\bibnamefont {Strunk}},
  \bibinfo {author} {\bibfnamefont {J.-P.}\ \bibnamefont {Salvetat}}, \bibinfo
  {author} {\bibfnamefont {J.-M.}\ \bibnamefont {Bonard}}, \bibinfo {author}
  {\bibfnamefont {L.}~\bibnamefont {Forró}}, \bibinfo {author} {\bibfnamefont
  {T.}~\bibnamefont {Nussbaumer}},\ and\ \bibinfo {author} {\bibfnamefont
  {C.}~\bibnamefont {Schönenberger}},\ }\bibfield  {title} {\bibinfo {title}
  {Aharonov–{Bohm} oscillations in carbon nanotubes},\ }\href
  {https://doi.org/10.1038/17755} {\bibfield  {journal} {\bibinfo  {journal}
  {Nature}\ }\textbf {\bibinfo {volume} {397}},\ \bibinfo {pages} {673}
  (\bibinfo {year} {1999})}\BibitemShut {NoStop}%
\bibitem [{\citenamefont {Brown~Jr.}(1963)}]{brownjr.Micromagnetics1963}%
  \BibitemOpen
  \bibfield  {author} {\bibinfo {author} {\bibfnamefont {W.~F.}\ \bibnamefont
  {Brown~Jr.}},\ }\href@noop {} {\emph {\bibinfo {title} {Micromagnetics}}}\
  (\bibinfo  {publisher} {{John Wiley \& Sons Inc}},\ \bibinfo {year}
  {1963})\BibitemShut {NoStop}%
\bibitem [{\citenamefont {Gurevich}\ and\ \citenamefont
  {Melkov}(1996)}]{gurevichMagnetizationOscillationsWaves1996}%
  \BibitemOpen
  \bibfield  {author} {\bibinfo {author} {\bibfnamefont {A.~G. A.~G.}\
  \bibnamefont {Gurevich}}\ and\ \bibinfo {author} {\bibfnamefont {G.~A.
  G.~A.}\ \bibnamefont {Melkov}},\ }\href@noop {} {\emph {\bibinfo {title}
  {Magnetization Oscillations and Waves}}}\ (\bibinfo  {publisher} {{CRC
  Press}},\ \bibinfo {year} {1996})\BibitemShut {NoStop}%
\bibitem [{\citenamefont {Stancil}(2009)}]{stancilSpinWavesTheory2009}%
  \BibitemOpen
  \bibfield  {author} {\bibinfo {author} {\bibnamefont {Stancil}},\ }\href
  {https://doi.org/10.1007/978-0-387-77865-5} {\emph {\bibinfo {title} {Spin
  {{Waves}} - {{Theory}} and {{Applications}}}}}\ (\bibinfo  {publisher}
  {{Springer US}},\ \bibinfo {address} {{Boston, MA}},\ \bibinfo {year}
  {2009})\BibitemShut {NoStop}%
\bibitem [{\citenamefont {K{\"o}rber}\ \emph {et~al.}(2021)\citenamefont
  {K{\"o}rber}, \citenamefont {Quasebarth}, \citenamefont {Otto},\ and\
  \citenamefont {K{\'a}kay}}]{korberFiniteelementDynamicmatrixApproach2021}%
  \BibitemOpen
  \bibfield  {author} {\bibinfo {author} {\bibfnamefont {L.}~\bibnamefont
  {K{\"o}rber}}, \bibinfo {author} {\bibfnamefont {G.}~\bibnamefont
  {Quasebarth}}, \bibinfo {author} {\bibfnamefont {A.}~\bibnamefont {Otto}},\
  and\ \bibinfo {author} {\bibfnamefont {A.}~\bibnamefont {K{\'a}kay}},\
  }\bibfield  {title} {\bibinfo {title} {Finite-element dynamic-matrix approach
  for spin-wave dispersions in magnonic waveguides with arbitrary cross
  section},\ }\href {https://doi.org/10.1063/5.0054169} {\bibfield  {journal}
  {\bibinfo  {journal} {AIP Advances}\ }\textbf {\bibinfo {volume} {11}},\
  \bibinfo {pages} {095006} (\bibinfo {year} {2021})}\BibitemShut {NoStop}%
\bibitem [{\citenamefont {Zimmermann}\ \emph {et~al.}(2018)\citenamefont
  {Zimmermann}, \citenamefont {Meier}, \citenamefont {Dirnberger},
  \citenamefont {K{\'a}kay}, \citenamefont {Decker}, \citenamefont {Wintz},
  \citenamefont {Finizio}, \citenamefont {Josten}, \citenamefont {Raabe},
  \citenamefont {Kronseder}, \citenamefont {Bougeard}, \citenamefont
  {Lindner},\ and\ \citenamefont
  {Back}}]{zimmermannOriginManipulationStable2018}%
  \BibitemOpen
  \bibfield  {author} {\bibinfo {author} {\bibfnamefont {M.}~\bibnamefont
  {Zimmermann}}, \bibinfo {author} {\bibfnamefont {T.~N.~G.}\ \bibnamefont
  {Meier}}, \bibinfo {author} {\bibfnamefont {F.}~\bibnamefont {Dirnberger}},
  \bibinfo {author} {\bibfnamefont {A.}~\bibnamefont {K{\'a}kay}}, \bibinfo
  {author} {\bibfnamefont {M.}~\bibnamefont {Decker}}, \bibinfo {author}
  {\bibfnamefont {S.}~\bibnamefont {Wintz}}, \bibinfo {author} {\bibfnamefont
  {S.}~\bibnamefont {Finizio}}, \bibinfo {author} {\bibfnamefont
  {E.}~\bibnamefont {Josten}}, \bibinfo {author} {\bibfnamefont
  {J.}~\bibnamefont {Raabe}}, \bibinfo {author} {\bibfnamefont
  {M.}~\bibnamefont {Kronseder}}, \bibinfo {author} {\bibfnamefont
  {D.}~\bibnamefont {Bougeard}}, \bibinfo {author} {\bibfnamefont
  {J.}~\bibnamefont {Lindner}},\ and\ \bibinfo {author} {\bibfnamefont {C.~H.}\
  \bibnamefont {Back}},\ }\bibfield  {title} {\bibinfo {title} {Origin and
  {{Manipulation}} of {{Stable Vortex Ground States}} in {{Permalloy
  Nanotubes}}},\ }\href {https://doi.org/10.1021/acs.nanolett.7b05222}
  {\bibfield  {journal} {\bibinfo  {journal} {Nano Letters}\ }\textbf {\bibinfo
  {volume} {18}},\ \bibinfo {pages} {2828} (\bibinfo {year}
  {2018})}\BibitemShut {NoStop}%
\bibitem [{\citenamefont {Server}()}]{mprime_m_m}%
  \BibitemOpen
  \bibfield  {author} {\bibinfo {author} {\bibfnamefont {B.~C.}\ \bibnamefont
  {Server}},\ }\href
  {https://www.cryst.ehu.es/cgi-bin/cryst/programs/corepresentations_point.pl?magnum=8.3.26}
  {\bibinfo {title} {{I}rreducible corepresentations of the {M}agnetic {P}oint
  {G}roup m'mm ({N}. 8.3.26)}}\BibitemShut {NoStop}%
\bibitem [{\citenamefont {Wagner}\ \emph {et~al.}(2016)\citenamefont {Wagner},
  \citenamefont {K{\'a}kay}, \citenamefont {Schultheiss}, \citenamefont
  {Henschke}, \citenamefont {Sebastian},\ and\ \citenamefont
  {Schultheiss}}]{wagnerMagneticDomainWalls2016}%
  \BibitemOpen
  \bibfield  {author} {\bibinfo {author} {\bibfnamefont {K.}~\bibnamefont
  {Wagner}}, \bibinfo {author} {\bibfnamefont {A.}~\bibnamefont {K{\'a}kay}},
  \bibinfo {author} {\bibfnamefont {K.}~\bibnamefont {Schultheiss}}, \bibinfo
  {author} {\bibfnamefont {A.}~\bibnamefont {Henschke}}, \bibinfo {author}
  {\bibfnamefont {T.}~\bibnamefont {Sebastian}},\ and\ \bibinfo {author}
  {\bibfnamefont {H.}~\bibnamefont {Schultheiss}},\ }\bibfield  {title}
  {\bibinfo {title} {Magnetic domain walls as reconfigurable spin-wave
  nanochannels},\ }\href {https://doi.org/10.1038/nnano.2015.339} {\bibfield
  {journal} {\bibinfo  {journal} {Nature Nanotechnology}\ }\textbf {\bibinfo
  {volume} {11}},\ \bibinfo {pages} {432} (\bibinfo {year} {2016})}\BibitemShut
  {NoStop}%
\bibitem [{\citenamefont {K{\"o}rber}\ \emph {et~al.}(2017)\citenamefont
  {K{\"o}rber}, \citenamefont {Wagner}, \citenamefont {K{\'a}kay},\ and\
  \citenamefont {Schultheiss}}]{korberSpinWaveReciprocityPresence2017}%
  \BibitemOpen
  \bibfield  {author} {\bibinfo {author} {\bibfnamefont {L.}~\bibnamefont
  {K{\"o}rber}}, \bibinfo {author} {\bibfnamefont {K.}~\bibnamefont {Wagner}},
  \bibinfo {author} {\bibfnamefont {A.}~\bibnamefont {K{\'a}kay}},\ and\
  \bibinfo {author} {\bibfnamefont {H.}~\bibnamefont {Schultheiss}},\
  }\bibfield  {title} {\bibinfo {title} {Spin-{{Wave Reciprocity}} in the
  {{Presence}} of {{N\'eel Walls}}},\ }\href
  {https://doi.org/10.1109/LMAG.2017.2762642} {\bibfield  {journal} {\bibinfo
  {journal} {IEEE Magnetics Letters}\ }\textbf {\bibinfo {volume} {8}},\
  \bibinfo {pages} {1} (\bibinfo {year} {2017})}\BibitemShut {NoStop}%
\bibitem [{\citenamefont {Ot{\'a}lora}\ \emph {et~al.}(2017)\citenamefont
  {Ot{\'a}lora}, \citenamefont {Yan}, \citenamefont {Schultheiss},
  \citenamefont {Hertel},\ and\ \citenamefont
  {K{\'a}kay}}]{otaloraAsymmetricSpinwaveDispersion2017}%
  \BibitemOpen
  \bibfield  {author} {\bibinfo {author} {\bibfnamefont {J.~A.}\ \bibnamefont
  {Ot{\'a}lora}}, \bibinfo {author} {\bibfnamefont {M.}~\bibnamefont {Yan}},
  \bibinfo {author} {\bibfnamefont {H.}~\bibnamefont {Schultheiss}}, \bibinfo
  {author} {\bibfnamefont {R.}~\bibnamefont {Hertel}},\ and\ \bibinfo {author}
  {\bibfnamefont {A.}~\bibnamefont {K{\'a}kay}},\ }\bibfield  {title} {\bibinfo
  {title} {Asymmetric spin-wave dispersion in ferromagnetic nanotubes induced
  by surface curvature},\ }\href {https://doi.org/10.1103/PhysRevB.95.184415}
  {\bibfield  {journal} {\bibinfo  {journal} {PHYSICAL REVIEW B}\ }\textbf
  {\bibinfo {volume} {95}},\ \bibinfo {pages} {184415} (\bibinfo {year}
  {2017})}\BibitemShut {NoStop}%
\bibitem [{\citenamefont {K{\"o}rber}\ and\ \citenamefont
  {K{\'a}kay}(2021)}]{korberNumericalReverseEngineering2021a}%
  \BibitemOpen
  \bibfield  {author} {\bibinfo {author} {\bibfnamefont {L.}~\bibnamefont
  {K{\"o}rber}}\ and\ \bibinfo {author} {\bibfnamefont {A.}~\bibnamefont
  {K{\'a}kay}},\ }\bibfield  {title} {\bibinfo {title} {Numerical reverse
  engineering of general spin-wave dispersions: Bridge between numerics and
  analytics using a dynamic-matrix approach},\ }\href
  {https://doi.org/10.1103/PhysRevB.104.174414} {\bibfield  {journal} {\bibinfo
   {journal} {Physical Review B}\ }\textbf {\bibinfo {volume} {104}},\ \bibinfo
  {pages} {174414} (\bibinfo {year} {2021})}\BibitemShut {NoStop}%
\bibitem [{\citenamefont {K\"orber}\ \emph {et~al.}(2021)\citenamefont
  {K\"orber}, \citenamefont {Zimmermann}, \citenamefont {Wintz}, \citenamefont
  {Finizio}, \citenamefont {Kronseder}, \citenamefont {Bougeard}, \citenamefont
  {Dirnberger}, \citenamefont {Weigand}, \citenamefont {Raabe}, \citenamefont
  {Ot\'alora}, \citenamefont {Schultheiss}, \citenamefont {Josten},
  \citenamefont {Lindner}, \citenamefont {K\'ezsm\'arki}, \citenamefont
  {Back},\ and\ \citenamefont {K\'akay}}]{korberSymmetryCurvatureEffects2021}%
  \BibitemOpen
  \bibfield  {author} {\bibinfo {author} {\bibfnamefont {L.}~\bibnamefont
  {K\"orber}}, \bibinfo {author} {\bibfnamefont {M.}~\bibnamefont
  {Zimmermann}}, \bibinfo {author} {\bibfnamefont {S.}~\bibnamefont {Wintz}},
  \bibinfo {author} {\bibfnamefont {S.}~\bibnamefont {Finizio}}, \bibinfo
  {author} {\bibfnamefont {M.}~\bibnamefont {Kronseder}}, \bibinfo {author}
  {\bibfnamefont {D.}~\bibnamefont {Bougeard}}, \bibinfo {author}
  {\bibfnamefont {F.}~\bibnamefont {Dirnberger}}, \bibinfo {author}
  {\bibfnamefont {M.}~\bibnamefont {Weigand}}, \bibinfo {author} {\bibfnamefont
  {J.}~\bibnamefont {Raabe}}, \bibinfo {author} {\bibfnamefont {J.~A.}\
  \bibnamefont {Ot\'alora}}, \bibinfo {author} {\bibfnamefont {H.}~\bibnamefont
  {Schultheiss}}, \bibinfo {author} {\bibfnamefont {E.}~\bibnamefont {Josten}},
  \bibinfo {author} {\bibfnamefont {J.}~\bibnamefont {Lindner}}, \bibinfo
  {author} {\bibfnamefont {I.}~\bibnamefont {K\'ezsm\'arki}}, \bibinfo {author}
  {\bibfnamefont {C.~H.}\ \bibnamefont {Back}},\ and\ \bibinfo {author}
  {\bibfnamefont {A.}~\bibnamefont {K\'akay}},\ }\bibfield  {title} {\bibinfo
  {title} {Symmetry and curvature effects on spin waves in vortex-state
  hexagonal nanotubes},\ }\href {https://doi.org/10.1103/PhysRevB.104.184429}
  {\bibfield  {journal} {\bibinfo  {journal} {Phys. Rev. B}\ }\textbf {\bibinfo
  {volume} {104}},\ \bibinfo {pages} {184429} (\bibinfo {year}
  {2021})}\BibitemShut {NoStop}%
\end{thebibliography}
\end{document}


{
\makeatletter
\def\frontmatter@thefootnote{%
 \altaffilletter@sw{\@fnsymbol}{\@fnsymbol}{\csname c@\@mpfn\endcsname}%
}%

\makeatother

\title{Supporting Information for "Mode splitting of spin waves in magnetic nanotubes with discrete symmetries"}

\author{Lukas K\"orber}\email{l.koerber@hzdr.de}
\affiliation{Helmholtz-Zentrum Dresden - Rossendorf, Institut f\"ur Ionenstrahlphysik und Materialforschung, D-01328 Dresden, Germany}
\affiliation{Fakultät Physik, Technische Universit\"at Dresden, D-01062 Dresden, Germany}

\author{Istv\'{a}n K\'{e}zsm\'{a}rki}
\affiliation{Experimental Physics V, University of Augsburg, 86135 Augsburg, Germany}

\author{Attila K\'{a}kay}
\affiliation{Helmholtz-Zentrum Dresden - Rossendorf, Institut f\"ur Ionenstrahlphysik und Materialforschung, D-01328 Dresden, Germany}

\date{\today}

\begin{abstract}
\end{abstract}

\maketitle

\section{S1: Numerical method}
Our numerical modeling is carried out within the continuum limit of micromagnetism \cite{brownjr.Micromagnetics1963, gurevichMagnetizationOscillationsWaves1996, stancilSpinWavesTheory2009}, which is adequate for spin waves with wave lengths in the nanometer regime and above. We obtain the spin-wave modes by solving the linearized torque equation of motion,
\begin{equation}\label{eq:dynamic-equation}
    \frac{\mathrm{d}}{\mathrm{d}t}(\delta \bm{m})=-\gamma\mu_0  \big[\delta \bm{m}\times \bm{H}_0 + \bm{m}_0\times \delta\bm{h}\big].
\end{equation}
which describes the magnetization dynamics $\delta\bm{m}(\bm{r},t)$ close to some equilibrium magnetization state $\bm{m}_0(\bm{r})$. Here, $\delta \bm{h}$ and $\bm{H}_0$ are the dynamic and static effective magnetic field acting on the magnetization. In our case, these fields are comprised by the exchange, dipole-dipole and uniaxial-anisotropy field. For the case of infinitely long tubular magnets where the equilibrium state is translationally invariant along the tube (here, $z$ direction), the solutions of Eq.~\eqref{eq:dynamic-equation} can be taken as $\bm{\eta}_{\nu,k} \exp[i(kz-\omega_\nu(k) t)]$ with the (complex) lateral spin-wave mode profiles $\bm{\eta}_{\nu,k}$, the wave vector in $z$ direction $k$, the angular frequency $\omega_\nu(k)$ and some lateral mode index $\nu$. Using this Ansatz, the equation of motion is transformed into an eigenvalue problem for the lateral mode profiles and the angular frequencies. To solve this eigenvalue problem numerically, we use a finite-element method to discretize the cross sections of the different polygonal nanotubes using small triangles. The details of this propagating-wave dynamic-matrix approach are found in \cite{korberFiniteelementDynamicmatrixApproach2021}. For this paper, we model infinitely-long regular-polygonal nanotubes with an outer diameter of \SI{250}{\nano\meter}, a thickness of $\SI{30}{\nano\meter}$, number of corners $c$ and typical material parameters for a soft ferromagnet. Namely, we assert a saturation magnetization of $M_\mathrm{s}=\SI{820}{\kilo\ampere/\meter}$, an exchange-stiffness constant of $A_\mathrm{ex}=\SI{13}{\pico\joule/\meter}$ and a reduced gyromagnetic ratio of $\gamma/2\pi = \SI{28}{\giga\hertz/\tesla}$. The average edge length of the triangles used in spatial discretization is approximately \SI{4}{\nano\meter} which is smaller than the characteristic exchange length of the material, under which the magnetization can be assumed to be homogeneous. Here, we model polygonal tubes between $c=6$ and $c=30$ corners, as well as the limiting case of a round tube, with $c=\infty$. The equilibrium state $\bm{m}_0$ is obtained for each tube by minimizing the total magnetic energy \cite{korberFiniteelementDynamicmatrixApproach2021}.

\section{S2: Irreducible representations of the corresponding magnetic point groups}

\begin{table}[htbp]
  \centering
  \caption{The irreducible representations and their dimensions of the magnetic point groups $c/m$'$m(m)$, to which the obtained spin-wave modes are categorized (or belong to). Here, $c$ is the number of facets of the nanotube. The symbol $E_*$ is a place holder for the different two-dimensional irreps which can exist depending on $c.$ For example, for a round tube ($c=\infty$), infinitely many 2D irreps exist. Also, in this case, the irrep $A_1$ contains only the homogeneous mode, $\nu=0$.}
  \begin{tabular}{*{6}{ccl|ccc}}
  \toprule
    irrep & dim. && $c=\infty$ & $c$ odd & $c$ even\\
    \midrule
        $E_*$ & 2 &
         doublets & 
        $\checkmark$ &
        $\checkmark$
          &
        $\checkmark$ \\
        $A_1$ & 1 & fully symmetric& (\checkmark)  &\checkmark &\checkmark \\
        $A_2$ & 1 & fully antisymmetric&  &\checkmark &\checkmark \\
        $\begin{array}{c} B_1 \\ B_2
        \end{array}$ & 
        $\begin{array}{c} 1 \\ 1
        \end{array}$ & $\Big\}$ mix of the two above\quad &
        &
        &
        $\begin{array}{c}
        \checkmark \\ \checkmark
        \end{array}$ \\
    \bottomrule
  \end{tabular}
  \label{tab:TAB1}
\end{table}

\section{S3: Evolution of the dispersion when lowering the rotational symmetry}

In the attached animated movie we show the transition of the dispersion relation when the number of facets of the polygonal magnetic tube is decreased from 30 to 6. We believe that the movie clearly shows the process of how the avoided level crossings \textit{follow} the branches, that are split by lowering the rotational symmetry, \textit{i.e.} they follow the singlet lines seen in Fig. 3(b) of the main manuscript.

%